\documentclass[12pt,titlepage]{article}

\usepackage{amsmath, amsfonts, natbib}
\usepackage[margin = 1in]{geometry}
\usepackage[parfill]{parskip}
\usepackage{setspace, placeins, graphicx, caption, subcaption, color, hyperref, tikz, authblk}
\usetikzlibrary{shapes,arrows,positioning}
\newtheorem{theorem}{Theorem}
\newtheorem{corollary}[theorem]{Corollary}
\newcommand\cites[1]{\citeauthor{#1}'s\ (\citeyear{#1})}

\def\keywords#1{{\vskip4pt
\noindent
\hbox to59.5pt{KEY\enspace WORDS:\quad\hss}\vtop{\advance \hsize by -59.5pt
\leftskip=28pt \rightskip=0pt
\noindent\ignorespaces#1\vskip8pt}}}
\title{On the relationship between set-based and network-based measures of gender homophily in scholarly publications\\ \quad \\
\large
Working Paper no. 157\\Center for Statistics and the Social Sciences\\
University of Washington \vspace{2cm}}

\author[1]{Y. Samuel Wang}
\author[1,2,3]{Elena A. Erosheva}
\affil[1]{Department of Statistics}
\affil[2]{School of Social Work}
\affil[3]{Center for Statistics and the Social Sciences}

\date{October 26, 2016}

\begin{document}
\pagenumbering{roman}

\maketitle

\begin{abstract}
There is an increased interest in the scientific community in the problem of measuring gender homophily in co-authorship on scholarly publications \citep{eisen:2016}. For a given set of publications and co-authorships, we assume that author identities have not been disambiguated in that we do not know when one person is an author on more than one paper. In this case, one way to think about measuring gender homophily is to consider all observed co-authorship pairs and obtain a set-based gender homophily coefficient \citep[e.g.,][]{berg:homophily2016}. Another way is to consider papers as observed disjoint networks of co-authors and use a network-based assortativity coefficient \citep[e.g.,][]{newman2003mixing}.
In this note, we review both metrics and show that the gender homophily set-based index is equivalent to the gender assortativity network-based coefficient with properly weighted edges. 

\end{abstract}

%\begin{footnotesize}
%\tableofcontents 
%\listoffigures
%\end{footnotesize}

\pagenumbering{arabic}
\baselineskip=18pt

\section{Introduction}
The phenomenon of individuals with similar characteristics more likely to form ties than individuals with dissimilar characteristics is known as assortativity or homophilly. When studying patterns in co-authorship in scientific publications, researchers typically consider sets of co-authorship occurrences and corresponding individual characteristics for some collection of papers, and measure homophily on a set of co-authorship occurrences. For example, previous studies in economics, an academic field dominated by men, have found evidence for gender-based homophily -- the principle that similarity breeds connection between individuals~\citep{mcpherson2001birds} -- in coauthorship. An earlier study that analyzed publications from a cohort sample of 178 PhDs in economics found that women were more than five times more likely than men to have women co-authors \citep{mcdowell1992effect}. A recent study that analyzed coauthor teams from 3,090 articles in the top three economics journals between 1991-2002 has found evidence in favor of gender-based homophily in team formation at the subfield level~\citep{boschini2007team}.
 
Given a set of co-authorship occurrences, \cite{bergstrom2003algebra} suggest using a coefficient of homophily $\alpha$ for set-based data where individuals take on a binary characteristic. The coefficient of homophily $\alpha$ has a simple and intuitive interpretation: the difference between the probability that a randomly chosen coauthor of a man is a man and the probability that a randomly chosen coauthor of a woman is a man. Bergstrom et al. (2016)\footnote{\url{http://eigenfactor.org/gender/assortativity/measuring\_homophily.pdf}} show that it is equal to the observed coauthor-gender correlation in the given collection of papers, and, in case of two-author papers, it is equal to Sewell Wright's coefficient of inbreeding \citep{wright1949genetical}.

In the presence of ties between individuals, another natural way to think about assortativity is through networks. Thus, assortative interactions have been studied in biological networks \citep{Piraveenan:2012:AMD:2077941.2077947}, networks among animals and fish \citep{LusseauS477,croft2005}, and social networks in humans \citep{foster2010edge,rivera2010dynamics}. Various metrics have been proposed to measure the assortativity within an observed network, including \cites{newman2003mixing} network-based assortativity coefficient where individuals are assigned a single categorical characteristic.
% while \cite{bergstrom2003algebra} proposes an index of assortativity $\alpha$ for set-based data where individuals take on a binary characteristic. In this report, we briefly review both metrics and show that in a network with ``appropriately weighted" edges, Newman's assortativity coefficient, $r$, is equivalent to Bergstrom's index of assortativity, $\alpha$.

In this paper, we consider a common scenario when gender indicators are known for coauthors on a set of publications but the author identities have not been disambiguated. We describe a set-based gender homophily coefficient \citep[e.g.,][]{berg:homophily2016} and show that it is equivalent to the network-based assortativity coefficient \citep[e.g.,][]{newman2003mixing} when edges within each paper are weighted inversely proportional to the number of co-authors on a paper.

\section{Newman's Measure of Assortativity}
We first consider a measure of assortativity defined by \citet{newman2003mixing} which explicitly assumes a network based representation. When the relational data is represented as a graph, each individual is represented as a node and edges between nodes indicate a relationship between the two nodes. If the relationship is asymmetric, the edges may be directed, or if the relationship is symmetric an undirected edge may be used. Assuming we have $i = 1, 2, \ldots K$ groups and that each individual in our sample belongs to a single group, let $e_{ij}$ be the proportion of all edges which point from an individual in category $i$ to an individual in category $j$, $a_i$ be the proportion of all edges which point from an individual in $i$ and $b_i$ be the proportion of all edges which point to an individual in category $i$. Then,
\begin{equation} \label{eq:rDef}
r = \frac{\sum_i e_{ii} - \sum_i a_ib_i}{1 - \sum_i a_ib_i}.
\end{equation}
When there is no observed assortativity, $r = 0$; when individuals form ties exclusively with other individuals with the same characteristic $r=1$; and when the network is perfectly dissasortive (each node is only connected to nodes of different characteristics) $r$ is negative and bounded below by -1 \citep{newman2003mixing}.

\begin{figure}[t]
\centering
  \begin{subfigure}[b]{.37\textwidth}
  \centering
      \begin{tikzpicture}[->,>=triangle 45,shorten >=1pt,
        auto,thick,
        main node/.style={circle,inner
          sep=6pt,draw,font=\sffamily}]  
      
        \node[main node] (1) [fill = gray]{}; 
        \node[main node] (2) [below=1cm of 1, fill = gray] {}; 
        \node[main node] (3) [below=1cm of 2, fill = gray] {}; 
        \node[main node] (4) [right=1.5cm of 1] {};
        \node[main node] (5) [right=1.5cm of 2] {};
        \node[main node] (6) [right=1.5cm of 3] {};

      \path[color=black!20!blue,every
              node/.style={font=\sffamily\small}] 
              [bend right=10](1) edge node {}(2)
              (2) edge node {}(1)
              (2) edge node {}(3)
              (3) edge node {}(2)
              [bend right = 45](1) edge node {}(3)
              [bend left = 30] (3) edge node {}(1)
              
              [bend left=10](4) edge node {}(5)
              (5) edge node {}(4)
              (5) edge node {}(6)
              (6) edge node {}(5)
              [bend left = 45](4) edge node {}(6)
              [bend right = 30] (6) edge node {}(4)
              [bend right = 20](1) edge node {}(4)
              [bend right = 20] (4) edge node {}(1)
              
              ;

      \end{tikzpicture}
      \caption{Assortative network; $r = .71$ and $\alpha = .78$}
      \end{subfigure}
      ~
        \begin{subfigure}[b]{.37\textwidth}
        \centering
            \begin{tikzpicture}[->,>=triangle 45,shorten >=1pt,
              auto,thick,
              main node/.style={circle,inner
                sep=6pt,draw,font=\sffamily}]  
            
              \node[main node] (1) [fill = gray]{}; 
              \node[main node] (2) [below=1cm of 1, fill = gray] {}; 
              \node[main node] (3) [below=1cm of 2, fill = gray] {}; 
              \node[main node] (4) [right=1.5cm of 1] {};
              \node[main node] (5) [right=1.5cm of 2] {};
              \node[main node] (6) [right=1.5cm of 3] {};
      
            \path[color=black!20!blue,every
                    node/.style={font=\sffamily\small}] 
                    [bend right=10](1) edge node {}(4)
                    (4) edge node {}(1)
                    (2) edge node {}(5)
                    (5) edge node {}(2)
                    (3) edge node {}(6)
                    (6) edge node {}(3)
                     [bend right = 45](1) edge node {}(3)
                     [bend left = 30] (3) edge node {}(1)
                                       
                    [bend left = 45](4) edge node {}(5)
                    [bend right = 30] (5) edge node {}(4)                    
                    ;

            \end{tikzpicture}
          
            \caption{Disassortative network; $r = -.13$ and $\alpha = -.33$}
            \end{subfigure}
\end{figure}

\FloatBarrier

\section{Bergstrom's $\alpha$}
For set-based data where individuals take on a binary characteristic, \cite{bergstrom2003algebra} proposes an index of assortativity $\alpha$  by constructing a difference in risks. More formally, suppose we have a single characteristic which is either positive of negative. Let $p$ be the probability that a randomly selected tie of a randomly selected positive individual connects to another positive individual. Let $q$ be the probability that a randomly selected tie of a randomly selected negative individual connects to a positive individual. We then define the assortativity measure as the difference of these risks $\alpha = p - q$. Because $\alpha$ is the difference between two probabilities, it must lie in [-1, 1]. The lower bound of -1 is only achieved in the extreme case where every individual has exactly one tie to an individual of the opposite characteristic, and the upper bound is only achieved in the extreme case where individuals form ties exclusively with others with the same characteristic.

Define $N_+$ and $N_-$ to be the set of all individuals with the positive and negative characteristics each of size $|N_+|$ and $|N_-|$ respectively.  Let $\pi_s$ and $\nu_s$ be the number of ties to positive and negative individuals for individual $s$. Finally let $K^\star$ be the size of the largest clique in the network (by assumption this is also equivalent to the max degree of the network), $[K^\star]$ denote the set $\{0,1,...K\}$ and $n_{ij}$ denote the number of cliques with $i$ positive individuals and $j$ negative individuals. We can then calculate $\alpha$ for a given network-
\begin{equation}
\setlength{\jot}{8pt}
\begin{split}
\alpha &= \frac{1}{|N_+|}\sum_{s \in N_+}\frac{\pi_s}{\pi_s + \nu_s} - \frac{1}{|N_-|}\sum_{s \in N_-}\frac{\pi_s}{\pi_s + \nu_s} \\
 & = \frac{1}{|N_+|}\sum_{i, j \in [K^\star] \times [K^\star]} n_{ij}\frac{i -1}{i + j - 1} -  \frac{1}{|N_-|}\sum_{i, j \in [K^\star] \times [K^\star]} n_{ij}\frac{i}{i + j - 1}.
 \end{split}  
\end{equation}
The first formulation arises explicitly from the difference of risks interpretation of $\alpha$, while the second formulation provides a computationally convenient way to calculate $\alpha$ from the sufficient statistics $n_{ij}$.

\section{Equivalence of $\alpha$ and $r$}
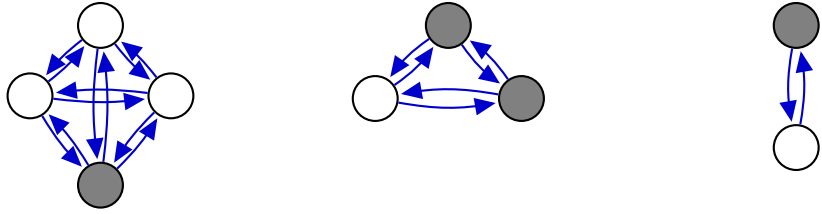
\begin{figure}
\centering
            \begin{tikzpicture}[->,>=triangle 45,shorten >=1pt,
              auto,thick,
              main node/.style={circle,inner
                sep=6pt,draw,font=\sffamily}]  
            
              \node[main node] (1) [fill = gray]{}; 
              \node[main node] (2) [below right=.75cm of 1, fill = gray] {}; 
              \node[main node] (3) [below left =.75cm of 1] {}; 
              
              \node[main node] (4) [right=4cm of 1, fill = gray] {};
              \node[main node] (5) [below=1cm of 4] {};

            \node[main node] (6) [left=4cm of 1] {};
            \node[main node] (7) [below=1.5cm of 6, fill = gray] {};
            \node[main node] (8) [below right =.7cm of 6] {};
            \node[main node] (9) [below left =.7cm of 6] {};

            \path[color=black!20!blue,every
                    node/.style={font=\sffamily\small}, bend right = 10] 
            (1) edge node {}(2)
            (2) edge node {}(1)
            (3) edge node {}(1)
            (1) edge node {}(3)
            (3) edge node {}(2)
            (2) edge node {}(3)
            
            (4) edge node {}(5)
			(5) edge node {}(4)

			[bend right = 7](6) edge node {}(7)
            (6) edge node {}(8)
            (6) edge node {}(9)
            (7) edge node {}(6)
            (7) edge node {}(8)
            (7) edge node {}(9)
            (8) edge node {}(6)
			(8) edge node {}(7)
            (8) edge node {}(9)
            (9) edge node {}(6)
            (9) edge node {}(7)
            (9) edge node {}(8)
              
      ;

            \end{tikzpicture}
\caption{\label{fig:cliques} Network of disjoint cliques. Each clique might represent an article and edges represent co-authorships. Note that we assume each individual has at least edge.}
\end{figure}

Because Newman's $r$ is a function of edge counts, individuals with higher degrees (number of edges) will influence the calculation of $r$ more than individuals with fewer co-authors. On the other hand, $\alpha$ explicitly places equal weight on each individual. Thus, although $r$ and $\alpha$ both measure assortativity, they are not equivalent in general. However, it can be shown that in a carefully specified network representation, Newman's $r$ is equal to $\alpha$. This work is motivated by the study of gender assortativity within co-authorships. In this case, the authors on each paper form a clique and the entire network is composed of disjoint cliques (a subset of nodes in which every node is connected to every other node; see Figure \ref{fig:cliques}). However, this result holds for any graph in which all edges are reciprocated (an edge from $s \rightarrow t $ implies there is also an edge from $t \rightarrow s$). 

Specifically, we construct a network $\mathcal{G} = \{V, E\}$, where $V$ is the set of all individuals and $E \subseteq V \times V$ denote the set of directed edges. If two individuals are tied, there are two directed edges, so that when individuals $s$ and $t$ are connected, both $s \rightarrow t$ and $s \leftarrow t$ are in $E$.

\begin{theorem}\label{thm:main}
In the graph $\mathcal{G} = \{V, E \}$, if each outgoing edge is weighted inversely to the node degree, then Newman's $r$ is equal to Bergstrom's $\alpha$
\end{theorem}

Intuitively, we can see that this is true because down weighting the edges of authors with many co-authors results in each author being counted equally regardless of the number of co-authors. 

The following corollary is a direct result of Theorem \ref{thm:main}.

\begin{corollary} \label{cor:equalAuth}
If every clique has the same number of individuals, then in the graph $\mathcal{G} = \{V, E\}$ where each edge has weight 1 Newman's $r$ is equal to Bergstrom's $\alpha$
\end{corollary}

\section{Proof of Theorem \ref{thm:main}}
Let $V$ be the set of individuals, and $N_+$ and $N_-$ be the set of positive and negative individuals respectively. Let $\pi_s$ be the number of edges from individual $s$ to a positive individual and $\nu_s$ be the number of edges from individual $s$ to a negative individual. Let $K_s = \pi_s + \nu_s$ denote the out-degree for individual $s$. Finally, let $Z_s$ denote the weight of edges for node $s$. 

Note that $e_{+-} = e_{-+}$ since for every edge from a positive to a negative individual, there must be the corresponding edge back form the negative to the positive. These quantities can be organized in a joint distribution table (same as Table 1 in Newman).

\begin{equation}
\begin{array}{|c | c | c | c|}
\hline
& + & - & Marginal \\ \hline
+ & e_{++} & e_{+-} & a_+ = e_{++} + e_{+-} \\ \hline
- & e_{-+} & e_{--} & a_- = e_{-+} + e_{--} \\ \hline
Marginal & b_+ = e_{++} + e_{-+} & b_- = e_{+-} + e_{--} & \\ \hline
\end{array}
\end{equation}

where the marginal quantities a and b are simply the row and column sums respectively. Note that since $e_{+-} = e_{-+}$, $a_i = b_i$.

Because we have only two groups and since the table is symmetric, then we have
\begin{equation} \label{eq:specificR}
\begin{split}
r &= \frac{\sum_i e_{ii} - \sum_i a_ib_i}{1 - \sum_i a_ib_i} \\
& = \frac{e_{++} + e_{--} - a_+^2 - a_-^2}{1 - a_+^2 - a_-^2} \\
\end{split}
\end{equation}

We then define the following quantities:
\begin{equation}
\begin{aligned}
\text{Weighted sum of } + \rightarrow + \text{ edges }&= \sum_{s \in N_+} Z_s\pi_s\\
\text{Weighted sum of } - \rightarrow - \text{ edges }& = \sum_{s \in N_-} Z_s\nu_s \\
\text{Weighted sum of } + \rightarrow - \text{ edges }& = \sum_{s \in N_+} Z_s\nu_s \\
\text{Weighted sum of } + \text{outgoing edges }& = \sum_{s \in N_+} Z_s(\pi_ s + \nu_s)  \\
\text{Weighted sum of } - \text{outgoing edges }& = \sum_{s \in N_-} Z_s(\pi_ s + \nu_s)  \\
\end{aligned}
\end{equation}
where the weighted proportion of the edges $e_{++}, e_{+-}, e_{--}, a_{+}, a_{-}$ are the quantities above normalized by the total weight of all edges
\[\sum_{s \in N_{+}} Z_s \pi_s + \sum_{s \in N_{+}}Z_s \nu_s + \sum_{s \in N_{-}} Z_s \pi_s + \sum_{s \in N_{-}}Z_s \nu_s = \sum_{s \in N_{+}} Z_s \pi_s + 2\sum_{s \in N_{+}}Z_s \nu_s + \sum_{s \in N_{-}}Z_s \nu_s\]
The simplification in the total weight uses the assumption that each reciprocated so $\sum_{s \in N_+} Z_s\nu_s = \sum_{s \in N_-} Z_s\pi_s$. 

First we simplify the numerator of $r$ from equation \ref{eq:specificR}-
\begin{equation}\scriptsize
\begin{aligned}
&\left(\sum_{s \in N_+} Z_s\pi_s + 2\sum_{s \in N_+} Z_s\nu_s +  \sum_{s \in N_-} Z_s\nu_s\right)^2e_{++} + e_{--} - a_+^2 - a_-^2 \\
&=\left(\sum_{s \in N_+} Z_s\pi_s + 2\sum_{s \in N_+}Z_s\nu_s + \sum_{s \in N_-}Z_s\nu_s\right)\left(\sum_{s \in N_+} Z_s\pi_s + \sum_{s \in N_-}Z_s\nu_s \right) + \left(\sum_{s \in N_+} Z_s\pi_s + \sum_{s \in N_+}Z_s\nu_s\right)^2 - \left(\sum_{s \in N_-}Z_s\nu_s + \sum_{s \in N_+}Z_s\nu_s\right)^2 \\
&=
\left(\sum_{s \in N_+} Z_s \pi_s\right)^2 + 2\left(\sum_{s \in N_+}Z_s\nu_s\right)\left(\sum_{s \in N_+} Z_s\pi_s\right) +   2\left(\sum_{s \in N_-}Z_s\nu_s\right)\left(\sum_{s \in N_+} Z_s\pi_s\right) + 2\left(\sum_{s \in N_+}Z_s\nu_s\right)\left(\sum_{s \in N_-}Z_s\nu_s\right) + \left(\sum_{s \in N_-}Z_s\nu_s\right)^2\\
&\quad 
- \left(\sum_{s \in N_+} Z_s\pi_s\right)^2 -  2\left(\sum_{s \in N_+} Z_s\pi_s\right)\left(\sum_{s \in N_+}Z_s\nu_s\right) - \left(\sum_{s \in N_+}Z_s\nu_s\right)^2 -\left(\sum_{s \in N_-}Z_s\nu_s\right)^2 -2\left(\sum_{s \in N_+}Z_s\nu_s\right)\left(\sum_{s \in N_-}Z_s\nu_s\right) - \left(\sum_{s \in N_+}Z_s\nu_s\right)^2\\
&= 
2 \left[ \left(\sum_{s \in N_-}Z_s\nu_s\right)\left(\sum_{s \in N_+} Z_s\pi_s\right) - \left(\sum_{s \in N_+}Z_s\nu_s\right)^2 \right]
\end{aligned}
\end{equation}

Now considering the denominator from equation \ref{eq:specificR},
\begin{equation}\scriptsize
\begin{aligned}
&\left(\sum_{s \in N_+} Z_s\pi_s + 2\sum_{s \in N_+} Z_s\nu_s +  \sum_{s \in N_-} Z_s\nu_s\right)^2\left(1 - a_1^2 - a_2^2\right) = \\
&\left(\sum_{s \in N_+} Z_s\pi_s + 2\sum_{s \in N_+}Z_s\nu_s + \sum_{s \in N_-}Z_s\nu_s\right)^2- 
\left(\sum_{s \in N_+} Z_s\pi_s + \sum_{s \in N_+}Z_s\nu_s\right)^2 - \left(\sum_{s \in N_-}Z_s\nu_s + \sum_{s \in N_+}Z_s\nu_s\right)^2 \\
&= 
\left(\sum_{s \in N_+} Z_s\pi_s\right)^2 + 4\left(\sum_{s \in N_+}Z_s\nu_s\right)^2 + \left(\sum_{s \in N_-}Z_s\nu_s\right)^2  +4\left(\sum_{s \in N_+}Z_s\nu_s\right)\left(\sum_{s \in N_+} Z_s\pi_s\right) + 4 \left(\sum_{s \in N_+}Z_s\nu_s\right)\left(\sum_{s \in N_-}Z_s\nu_s\right)\\
& \quad + 2\left(\sum_{s \in N_-}Z_s\nu_s\right)\left(\sum_{s \in N_+} Z_s\pi_s\right)- \left(\sum_{s \in N_+}Z_s\pi_s\right)^2  - 2\left(\sum_{s \in N_+} Z_s\pi_s\right)\left(\sum_{s \in N_+}Z_s\nu_s\right) - \left(\sum_{s \in N_+}Z_s\nu_s\right)^2  -\left(\sum_{s \in N_-}Z_s\nu_s\right)^2\\
& \quad  - 2\left(\sum_{s \in N_-}Z_s\nu_s\right)\left(\sum_{s \in N_+}Z_s\nu_s\right) - \left(\sum_{s \in N_+}Z_s\nu_s\right)^2 \\
& =
2\left(\sum_{s \in N_+}Z_s\nu_s\right)^2 + 2\left(\sum_{s \in N_+} Z_s\pi_s\right)\left(\sum_{s \in N_+}Z_s\nu_s\right) \\
&\quad + 2 \left(\sum_{s \in N_+}Z_s\nu_s\right)\left(\sum_{s \in N_-}Z_s\nu_s\right) + 2 \left(\sum_{s \in N_-}Z_s\nu_s\right)\left(\sum_{s \in N_+} Z_s\pi_s\right)\\
& = 
2\left(\sum_{s \in N_+}Z_s\nu_s\right)\left(\left(\sum_{s \in N_+}Z_s\nu_s\right) + \left(\sum_{s \in N_+}Z_s\pi_s\right)\right) +  2\left(\sum_{s \in N_-}Z_s\nu_s\right)\left(\left(\sum_{s \in N_+}\nu_s\right) + \left(\sum_{s \in N_+}Z_s\pi_s\right)\right) \\
& = 2
\left(\sum_{s \in N_+}Z_s\nu_s + \sum_{s \in N_-}Z_s\nu_s\right)\left(\sum_{s \in N_+}Z_s\nu_s + \sum_{s \in N_+}Z_s\pi_s\right)
\end{aligned}
\end{equation}

Recall that $\sum_{s\in N_-}\pi_s = \sum_{s\in N_+}\nu_s$. Simplifying the numerator and denominator together yields-
\begin{equation}\scriptsize
\begin{aligned}
r &= \frac{ \left(\sum_{s \in N_-} Z_s\nu_s \right)\left(\sum_{s \in N_+} Z_s\pi_s \right) - \left(\sum_{s \in N_+} Z_s\nu_s \right)^2}{\left(\sum_{s \in N_+}Z_s\pi_s + \sum_{s \in N_+}  Z_s\nu_s\right)\left(\sum_{s \in N_-} Z_s\nu_s + \sum_{s \in N_+} Z_s\nu_s \right) }\\
& = \frac{ \left(\sum_{s \in N_+} Z_s\pi_s\right) \left(\sum_{s \in N_-} Z_s\nu_s + \sum_{s \in N_+} Z_s\nu_s \right) + \left(\sum_{s \in N_-} Z_s\nu_s \right) \left(\sum_{s \in N_+}Z_s\pi_s + \sum_{s \in N_+}  Z_s\nu_s \right) }{\left(\sum_{s \in N_+}Z_s\pi_s + \sum_{s \in N_+}  Z_s\nu_s\right)\left(\sum_{s \in N_-} Z_s\nu_s + \sum_{s \in N_+} Z_s\nu_s \right)} \\
&\quad - \frac{\left(\sum_{s \in N_+}Z_s\pi_s + \sum_{s \in N_+}  Z_s\nu_s\right)\left(\sum_{s \in N_-} Z_s\nu_s + \sum_{s \in N_+} Z_s\nu_s \right)}{ \left(\sum_{s \in N_+}Z_s\pi_s + \sum_{s \in N_+}  Z_s\nu_s\right)\left(\sum_{s \in N_-} Z_s\nu_s + \sum_{s \in N_+} Z_s\nu_s \right)} \\
& = \frac{\sum_{s \in N_+} Z_s\pi_s}{\sum_{s \in N_+}Z_s\pi_s + \sum_{s \in N_+} Z_s \nu_s} + \frac{\sum_{s \in N_-}Z_s \nu_s }{\sum_{s \in N_-} Z_s\nu_s + \sum_{s \in N_+} Z_s\nu_s}  - 1\\
& = \frac{\sum_{s \in N_+} Z_s\pi_s}{\sum_{s \in N_+}Z_s\pi_s + \sum_{s \in N_+}  Z_s\nu_s} - \frac{\sum_{s \in N_-} Z_s\pi_s }{\sum_{s \in N_+} Z_s\nu_s + \sum_{s \in N_-} Z_s\nu_s}\\
\end{aligned}
\end{equation}

When $Z_s = \frac{c}{K_s}$ for some constant $c$, the denominators simplify
\begin{equation}
\sum_{s \in N_+}Z_s\pi_s + \sum_{s \in N_+}  Z_s\nu_s = \sum_{s \in N_+}\frac{c}{K_s} \left(\pi_s + \nu_s\right) = \sum_{s \in N_+} c = c|N_+|
\end{equation}

\begin{equation}
\sum_{s \in N_-}Z_s\nu_s + \sum_{s \in N_+}  Z_s\nu_s = \sum_{s \in N_-}Z_s\nu_s + \sum_{s \in N_-}  Z_s\pi_s = \sum_{s \in N_-}\frac{c}{K_s} \left(\pi_s + \nu_s\right) = \sum_{s \in N_-} c = c|N_-|
\end{equation}
and
\begin{equation}
r = \frac{\sum_{s \in N_+} \frac{c}{K_s}\pi_s}{c|N_+|} -\frac{\sum_{s \in N_-}\frac{c}{K_s} \pi_s }{c|N_-|} = \frac{1}{|N_+|}\sum_{s \in N_+}\frac{ \pi_s}{K_s} - \frac{1}{|N_-|}\sum_{s \in N_-}\frac{ \pi_s}{K_s} = \alpha
\end{equation}

Thus, Newman's assortativity coefficient, when obtained from a network of co-authorships where edges are weighted inversely proportional to the number of co-authors, is equal to $\alpha$. 

For a proof of Corollary \ref{cor:equalAuth}, if all authors have the same number of co-authors K, then we let c = K so that $Z_s = \frac{c}{K_s} = \frac{K}{K} = 1$ and all edges have weight 1.

\section{Conclusion}
Scientists have become increasingly aware of the gender imbalances present in professional academic activities \citep[e.g.,][]{west2013role}, and have raised the issue of importance of proper analysis and measurement \citep{eisen:2016}. In this note, by showing under what circumstances Bergstrom's $\alpha$ and Newman's $r$ are equivalent, we also hope to highlight how they differ. In particular, we note that authors with many co-authors (or equivalently papers with many authors) have a greater effect on the originally proposed Newman's $r$ (with all edge weights equal) than they would have in Bergstrom's $\alpha$. This may or may not be desirable and should be considered carefully depending on the specific context.

\bibliography{aeBib}

\end{document}